\documentclass[twocolumn,aps,prc,superscriptaddress,showpacs,floatfix,longbibliography]{revtex4-1}
\usepackage{url}
\usepackage{cancel}
\usepackage[colorlinks,linkcolor=blue,citecolor=blue,filecolor=black,urlcolor=blue]{hyperref}
\usepackage{epsfig,graphics}
\usepackage{graphicx}
\usepackage{dcolumn}
\usepackage{bm}
\usepackage[usenames]{color}
\usepackage{amssymb}
\usepackage{amsmath}
\usepackage{multirow}
\usepackage{float}
\usepackage{harpoon}
\usepackage{MnSymbol}
\usepackage{appendix}
\usepackage{color}
\usepackage{hyperref}
\usepackage{cleveref}

\newcommand{\sqrtsnn}{\mbox{$\sqrt{s^{}_{\mathrm{NN}}}$}}

\newcommand{\auau}{$^{197}$Au+$^{197}$Au}

\newcommand{\nene}{$^{20}$Ne+$^{20}$Ne}

\begin{document}
\title{Directly probing existence of $\alpha$-cluster structure in $^{20}$Ne by relativistic heavy-ion collisions}
\author{Lu-Meng Liu}
\affiliation{Physics Department and Center for Particle Physics and Field Theory, Fudan University, Shanghai 200438, China}

\author{Hai-Cheng Wang}\email[Hai-Cheng Wang and Lu-Meng Liu contributed equally to this work.]{}
\affiliation{School of Physics Science and Engineering, Tongji University, Shanghai 200092, China}

\author{Song-Jie Li}
\affiliation{School of Physics Science and Engineering, Tongji University, Shanghai 200092, China}

\author{Chunjian Zhang}
\affiliation{Key Laboratory of Nuclear Physics and Ion-beam Application (MOE), Fudan University, Shanghai 200433, China}
\affiliation{Shanghai Research Center for Theoretical Nuclear Physics, National Natural Science Foundation of China and Fudan University, Shanghai 200438, China}

\author{Jun Xu}\email{junxu@tongji.edu.cn}
\affiliation{School of Physics Science and Engineering, Tongji University, Shanghai 200092, China}

\author{Zhong-Zhou Ren}\email{zren@tongji.edu.cn}
\affiliation{School of Physics Science and Engineering, Tongji University, Shanghai 200092, China}

\author{Jiangyong Jia}\email{jiangyong.jia@stonybrook.edu}
\affiliation{Department of Chemistry, Stony Brook University, Stony Brook, NY 11794, USA}
\affiliation{Physics Department, Brookhaven National Laboratory, Upton, NY 11976, USA}

\author{Xu-Guang Huang}\email{huangxuguang@fudan.edu.cn}
\affiliation{Physics Department and Center for Particle Physics and Field Theory, Fudan University, Shanghai 200438, China}
\affiliation{Key Laboratory of Nuclear Physics and Ion-beam Application (MOE), Fudan University, Shanghai 200433, China}
\affiliation{Shanghai Research Center for Theoretical Nuclear Physics, National Natural Science Foundation of China and Fudan University, Shanghai 200438, China}

\date{\today}
\begin{abstract}
Can relativistic heavy-ion collisions only probe the global shape of colliding nuclei, or their detailed internal structure as well? Taking $^{20}$Ne as an example, we attempt to directly probe its internal $\alpha$-cluster structure, by comparing experimentally measured observables in collisions at relativistic energies from density distributions of $^{20}$Ne with and without $\alpha$-cluster structure. Since the two density distributions give the same nucleus size and deformation, they lead to similar mid-rapidity observables. However, the $\alpha$-cluster structure may considerably reduce the free spectator nucleon yield and enhance the spectator light nuclei yield, as a result of more compact initial phase-space distribution of nucleons inside $\alpha$ clusters. We propose to measure the scaled yield ratio of free spectator neutrons to charged particles with mass-to-charge ratio $A/Z = 3$, 3/2, and 2 in ultra-central \nene\ collisions, which is found to be reduced by about $25\%$ at \(\sqrt{s_\mathrm{NN}} = 7\) TeV and about 20\% at \(\sqrt{s_\mathrm{NN}} = 200\) GeV with $\alpha$-cluster structure in $^{20}$Ne. This scaled yield ratio thus serves as a robust and direct probe of the existence of $\alpha$-cluster structure in $^{20}$Ne free from the uncertainty of mid-rapidity dynamics.
\end{abstract}
\maketitle

\textbf{Introduction.}
Understanding the nucleus structure is a fundamental goal of nuclear physics, and recently it was proposed that relativistic heavy-ion collisions could be an alternative way of achieving this goal~\cite{Jia:2022ozr}. Typically, by analyzing probes such as the anisotropic flow, the fluctuation of mean transverse momentum, as well as their correlations~\cite{Jia:2021tzt,Jia:2021qyu}, people are able to extract the deformation of colliding nuclei, and this works successfully for heavy nuclei such as $^{197}$Au~\cite{Giacalone:2021udy}, $^{96}$Zr~\cite{Zhang:2021kxj}, $^{96}$Ru~\cite{Zhang:2021kxj}, and $^{238}$U~\cite{STAR:2024wgy}, whose density distributions can be well described by a deformed Woods-Saxon (WS) form. For light nuclei, such as $^{12}$C, $^{16}$O, and $^{20}$Ne, while they may have large deformation, various studies have shown that there could be $\alpha$-cluster structures in these nuclei~\cite{Freer:2017gip,BIJKER2020103735,Tohsaki:2017hen,Zhou:2023vgv,PhysRevLett.119.222505,Shen:2022bak}. Depending on the specific $\alpha$-cluster configuration, their density distributions may or may not be well described by a deformed WS form, and this may affect the effectiveness of the deformation probes as mentioned above. As shown in the recent study by some of us~\cite{Wang:2024ulq}, the density distributions of the energy-favored tetrahedron structure of $^{16}$O and the bowling-pin structure of $^{20}$Ne can be fairly fitted by a deformed WS form, and it is difficult to distinguish the real density distribution and the assumed deformed WS distribution by using mid-rapidity observables, e.g., the deformation probes mentioned above. Whether relativistic heavy-ion collisions can be used to probe only the global shape of colliding nuclei, or can also be used to probe directly the existence of internal $\alpha$-cluster structure in light nuclei, is a question to be answered.

Fortunately, besides mid-rapidity observables as mentioned above, particle yields at forward- and backward-rapidity in relativistic heavy-ion collisions from the multifragmentation process of the spectator matter may also provide valuable information of the structure of colliding nuclei~\cite{Liu:2022kvz,Liu:2022xlm,Liu:2023qeq,Liu:2023rap}. Free spectator neutrons, for instance, can be measured using zero-degree calorimeters (ZDC)~\cite{PHENIX:2000owy, ALICE:2013hur}, and free charged spectator particles can be measured with dedicated detectors positioned in the forward region~\cite{Tarafdar:2014oua}. These particles do not experience the complicated mid-rapidity dynamics and serve actually as more robust probes. In the previous study by some of us based on the Glauber model, we have shown that different $\alpha$-cluster structures in \(^{12}\)C and \(^{16}\)O may lead to different spectator particle yields in their collisions at relativistic energies~\cite{Liu:2023gun}. In the present study, we will investigate the possibility of disentangling the density distributions of $^{20}$Ne, which have a similar overall shape but with and without $\alpha$-cluster structure inside. By using central \nene\ collisions at relativistic energies, while one is difficult to probe the existence of $\alpha$-cluster structure with mid-rapidity observables, we will show that spectator particle yields, particularly the scaled yield ratios of free spectator neutrons to charged particles, may directly probe the existence of $\alpha$-cluster structure in $^{20}$Ne. Our study provides valuable references for heavy-ion collision experiments with light nuclei at both RHIC~\cite{huang2023measurementsazimuthalanisotropies16o16o} and LHC~\cite{brewer2021opportunitiesoopocollisions} energies.

\begin{figure}[ht]
\includegraphics[scale=0.15]{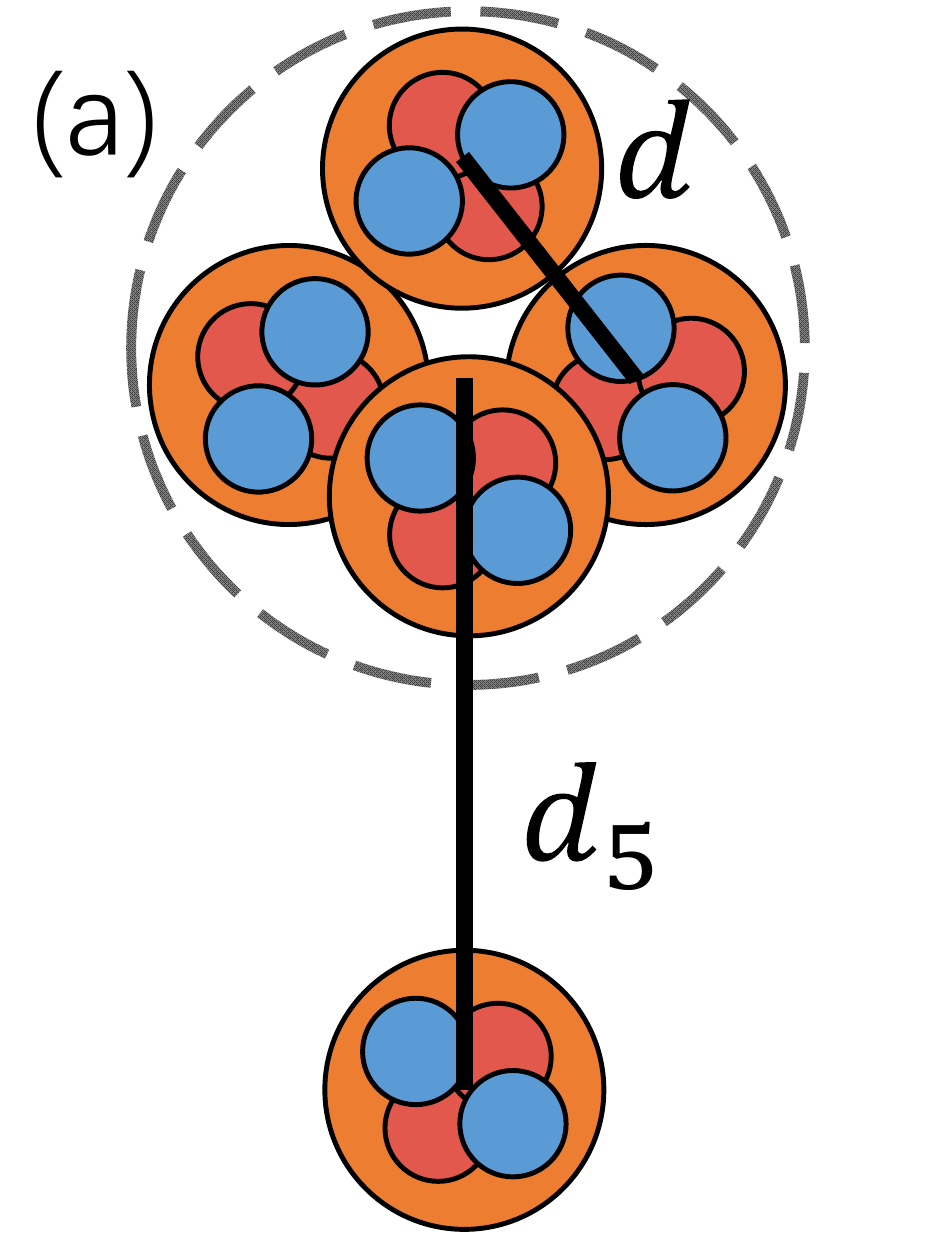}
\includegraphics[width=0.7\linewidth]{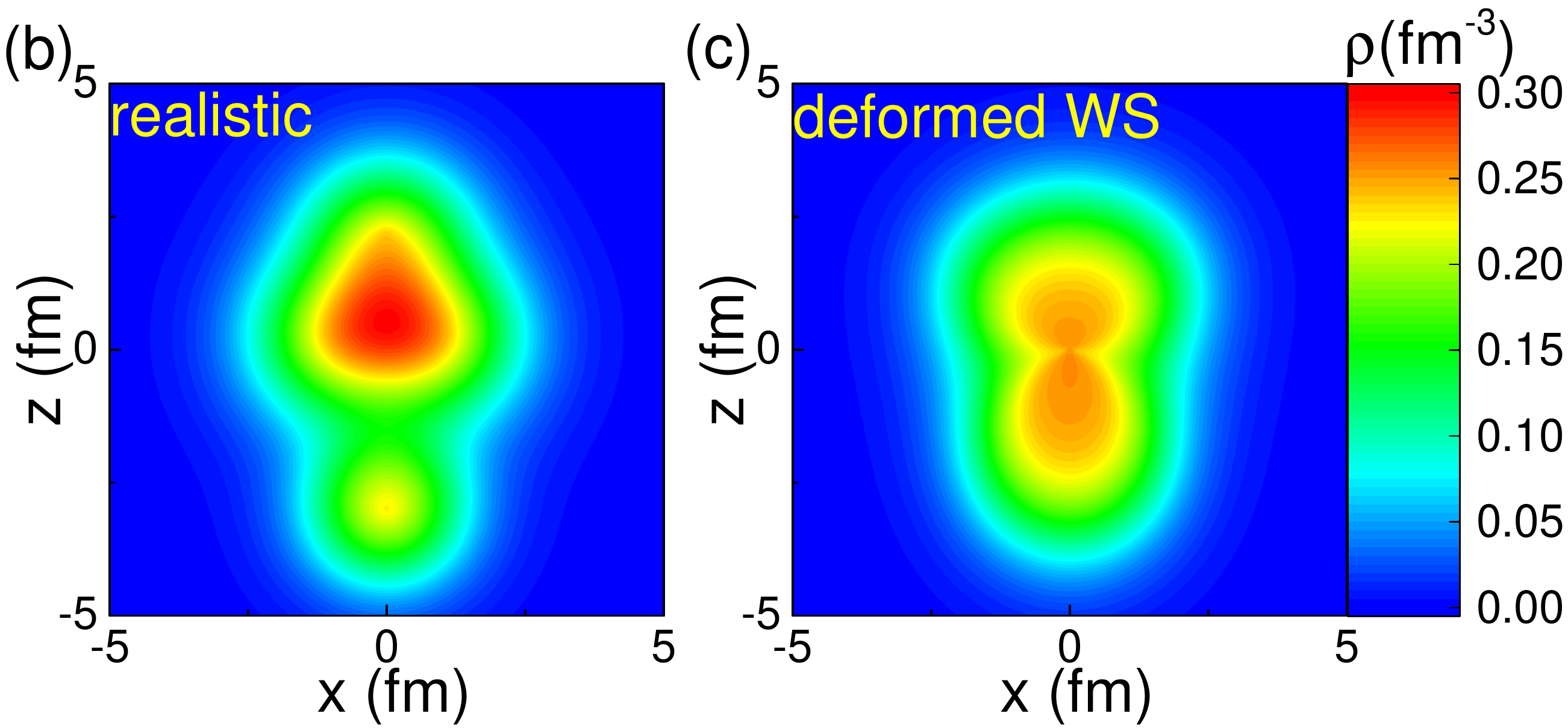}
\caption{(a): Bowling-pin configuration of $^{20}$Ne with five $\alpha$ clusters; (b): Realistic density distribution of $^{20}$Ne with $\alpha$ clusters; (c): Fitted density distribution of $^{20}$Ne with a deformed WS form.}\label{fig:densitydis}
\end{figure}

\textbf{Framework.}
The density distribution of $^{20}$Ne is obtained from a microscopic cluster model with Bloch-Brink wave function. The nucleon-nucleon effective interactions contain the Volkov No.2 force~\cite{Volkov_1965_NP} as well as the spin-orbit interaction. As shown in Fig.~\ref{fig:densitydis} (a), we assume that five $\alpha$ clusters form a bowling-pin structure in $^{20}$Ne, with four $\alpha$ clusters forming a tetrahedral configuration, and the fifth $\alpha$ cluster placed beneath the tetrahedron~\cite{Yamaguchi:2023mya}. Values of the distance parameters $d$ and $d_5$ are determined by minimizing the total energy after the angular-momentum projection of the total wave function. This leads to the realistic density distribution of $^{20}$Ne with $\alpha$-cluster structure, and its longitudinal section is shown in Fig.~\ref{fig:densitydis} (b). For details of the method to obtain the density distribution, we refer the reader to Ref.~\cite{Wang:2024ulq}. We have further fitted this density distribution with an axially symmetric deformed WS form
\begin{equation}\label{eq:WS}
\rho(r,\theta) = \frac{\rho_0}{1+\exp[(r-R(\theta))/a]},
\end{equation}
where $\rho_0$ is the normalization constant, and
\begin{equation}\label{eq:WSR}
R(\theta)=R_0[1+\sum_{n=2,3} \beta_n Y_{n,0}(\theta)]
\end{equation}
is the deformed radius, with $Y_{n,0}$ being the spherical harmonics. Values of the radius $R_0$, the diffuseness parameter $a$, and the deformation coefficients $\beta_{2,3}$ are determined by the moments \(\langle r^2 \rangle\) and \(\langle r^4 \rangle\) as well as the intrinsic multipole moments $Q_{2,3}$ obtained from the realistic density distribution of \(^{20}\)Ne as shown in Fig.~\ref{fig:densitydis} (b). The above moments are calculated from
\begin{eqnarray}
\langle r^l \rangle &=& \frac{\int \rho(\vec{r}) r^l \, d^3 r}{\int \rho(\vec{r}) \, d^3 r}, \\
Q_n &=& \int \rho(\vec{r}) r^n Y_{n,0}(\theta) \, d^3 r.
\end{eqnarray}
The longitudinal section of the resulting density distribution fitted by the deformed WS form is displayed in Fig.~\ref{fig:densitydis} (c). In this way, the two density distributions as shown in Fig.~\ref{fig:densitydis} (b) and (c) have the similar overall shape.

The study is based on an event-by-event hybrid simulation framework that combines a multiphase transport (AMPT) model~\cite{Lin:2004en} with the spectator particle generator (specCOAL+TestP+GEMINI)~\cite{Liu:2022xlm}, with the above density distribution at randomized collision orientation as the initial condition. Different from the original AMPT model, the initial momenta of all nucleons are sampled isotropically within an isospin-dependent Fermi sphere, with the Fermi momentum determined by the local density of neutrons or protons. Mid-rapidity observables are calculated from the final hadron phase-space information generated by the string melting AMPT model, with the setups the same as in Ref.~\cite{Wang:2024ulq}. Spectator nucleons are those that do not experience any elastic or inelastic collisions, and they are then grouped into heavy clusters (with $A \geq 4$) and free nucleons using a minimum spanning tree algorithm. Specifically, nucleons with a spatial distance $\Delta r < \Delta r_{\mathrm{max}}$ and relative momentum $\Delta p < p_{\mathrm{max}}$ are considered to form heavy clusters. The coalescence parameters $\Delta r_{\mathrm{max}} = 3$ fm and $\Delta p_{\mathrm{max}} = 300$ MeV/$c$, taken from Ref.~\cite{Li:1997rc}, have been shown to provide the best description of the experimental data for free spectator neutrons in ultra-central \auau\ collisions at $\sqrt{s_\mathrm{NN}} = 130$ GeV~\cite{Liu:2022kvz}. For spectator nucleons that do not form heavy clusters ($A \geq 4$), they may coalesce into light clusters ($A \leq 3$), such as deuterons, tritons, and $^3$He, based on a Wigner function approach~\cite{Chen:2003ava,Sun:2017ooe}. The deexcitation of heavy clusters with $A \geq 4$ is described by the GEMINI model~\cite{Charity:1988zz}, which reproduces the kinetic energy spectra of nucleons and $\alpha$ particles in low-energy reactions rather well~\cite{Charity:2010wk}. The deexcitation process in the GEMINI model depends the angular momentum and the excitation energy of the cluster. The angular momentum is computed by summing the contributions from all nucleons relative to the cluster's center of mass. The excitation energy is calculated by subtracting the ground-state energy (from either the mass table~\cite{Wang:2021xhn} or an improved liquid-drop model~\cite{Wang:2014qqa}) from the total cluster energy. We calculate the total cluster energy based on a simplified Skyrme-type energy-density functional~\cite{Chen:2010qx}, with the nucleon density distribution obtained from the test-particle method~\cite{Wong:1982zzb, Bertsch:1988ik} using the neutron and proton phase-space information of 200 parallel events for the same collision orientation. For details of the multifragmentation treatment of the spectator matter, we refer the reader to Ref.~\cite{Liu:2022xlm}. The AMPT+specCOAL+testP+GEMINI framework allows for investigating observables at both midrapidities and forward/backward rapidities in a simultaneous way.

\begin{figure}[ht]
\includegraphics[width=1.\linewidth]{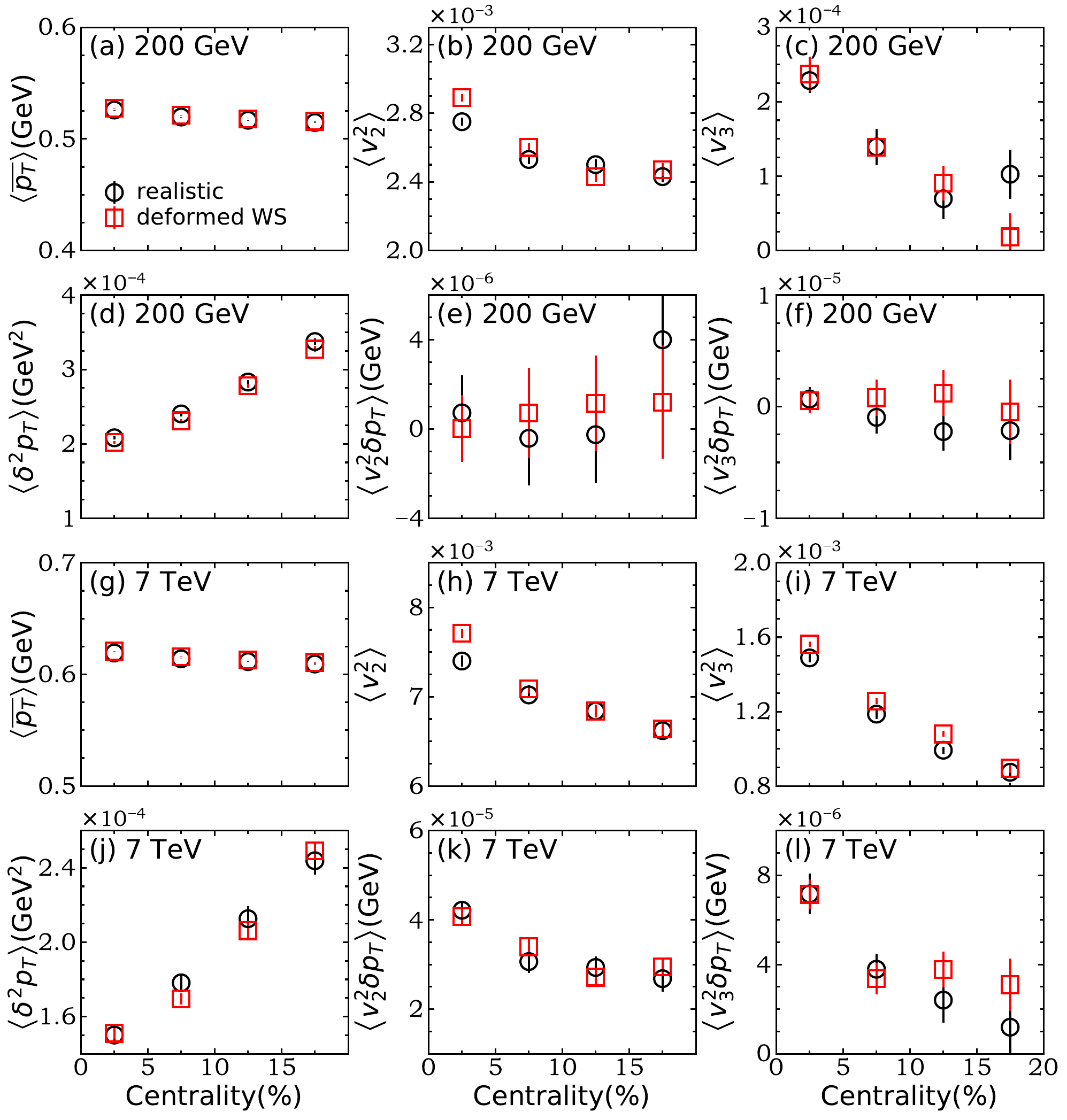}
\vspace*{-.5cm}
\caption{Mid-rapidity observables (\(\langle \overline{p_T} \rangle\), \(\langle v_n^2 \rangle\), \(\langle \delta p_T^2 \rangle\), and \(\langle v_n^2 \delta p_T \rangle\)) at \(0-20\%\) centrality in \nene\ collisions at $\sqrtsnn = 200$ GeV [(a)-(f)] and 7 TeV [(g)-(l)] from the realistic and deformed WS density distributions shown in Fig.~\ref{fig:densitydis}.}\label{fig:vndeltapt}
\end{figure}

\textbf{Mid-rapidity probes.}
To investigate the possibility of disentangling the density distributions shown in Fig.~\ref{fig:densitydis} (b) and (c), we begin by illustrating the resulting mid-rapidity observables in central \(^{20}\)Ne + \(^{20}\)Ne collisions at \(\sqrt{s_\mathrm{NN}} = 200\) GeV and 7 TeV in Fig.~\ref{fig:vndeltapt}. Here, the anisotropic flow $\langle v_n^2 \rangle$ and the fluctuation of the transverse momentum $\langle \delta p_T^2 \rangle$ as well as their correlation $\langle v_n^2 \delta p_T \rangle$ are calculated using the final hadron phase-space information from AMPT+specCOAL+testP+GEMINI according to
\begin{eqnarray}
\langle v_n^2 \rangle &=& \langle \cos [n(\varphi_i-\varphi_j)] \rangle_{i,j}, \\
\langle \delta p_T^2 \rangle &=& \langle (p_{T,i} - \langle \overline{p_T}\rangle) (p_{T,j} - \langle \overline{p_T}\rangle) \rangle_{i,j}, \\
\langle v_n^2 \delta p_T \rangle &=& \langle \cos[n(\varphi_i-\varphi_j)] (p_{T,k} - \langle \overline{p_T}\rangle)\rangle_{i,j,k}.
\end{eqnarray}
In the above, $p_{T,i}=\sqrt{p_{x,i}^2+p_{y,i}^2}$ and $\varphi_i=\arctan(p_{y,i}/p_{x,i})$ are, respectively, the transverse momentum and its azimuthal angle of the $i$th particle, $\langle \overline{p_T} \rangle$ is the mean transverse momentum, with $\overline{(...)}$ representing the average over all particles in one event. We take the convention that $\langle...\rangle$ represents the average over all events, and $\langle...\rangle_{i,j,...}$ represents the average over all possible combinations of $i,j,...$ for all events. Hadrons at midpseudorapidities ($|\eta|<2$) and $0.2<p_T<3$ GeV are selected for the calculation, with a pseudorapidity gap of $|\Delta \eta|>0.5$ used in evaluating $\langle...\rangle_{i,j,...}$ to remove the non-flow effect.

As shown in previous studies~\cite{Zhang:2021kxj,Giacalone:2021udy,STAR:2024wgy}, the selected mid-rapidity observables are probes of the area and anisotropy as well as their fluctuations and correlations in the initial overlap region. However, they are mostly not distinguishable within error bars from the two density distributions of $^{20}$Ne. The largest difference in $\langle v^2_2 \rangle$ is only about $6\%$ at 200 GeV and about $4\%$ at 7 TeV in ultra-central collisions (UCC). We have also tried to search for other mid-rapidity observables, such as the yield ratio of light nuclei~\cite{Sun:2017xrx} and net-proton fluctuation~\cite{Luo:2017faz}, etc., but ended up with the conclusion that it is difficult to use mid-rapidity observables to distinguish the difference in the two density distributions of $^{20}$Ne. This is not surprising, since these observables most manifest the global properties of the shape and energy density of the initial quark-gluon plasma, while effects of clusters due to nucleon-nucleon correlations are mostly smeared out by mid-rapidity dynamics.

\begin{figure}[ht]
\centering
\includegraphics[width=1.\linewidth]{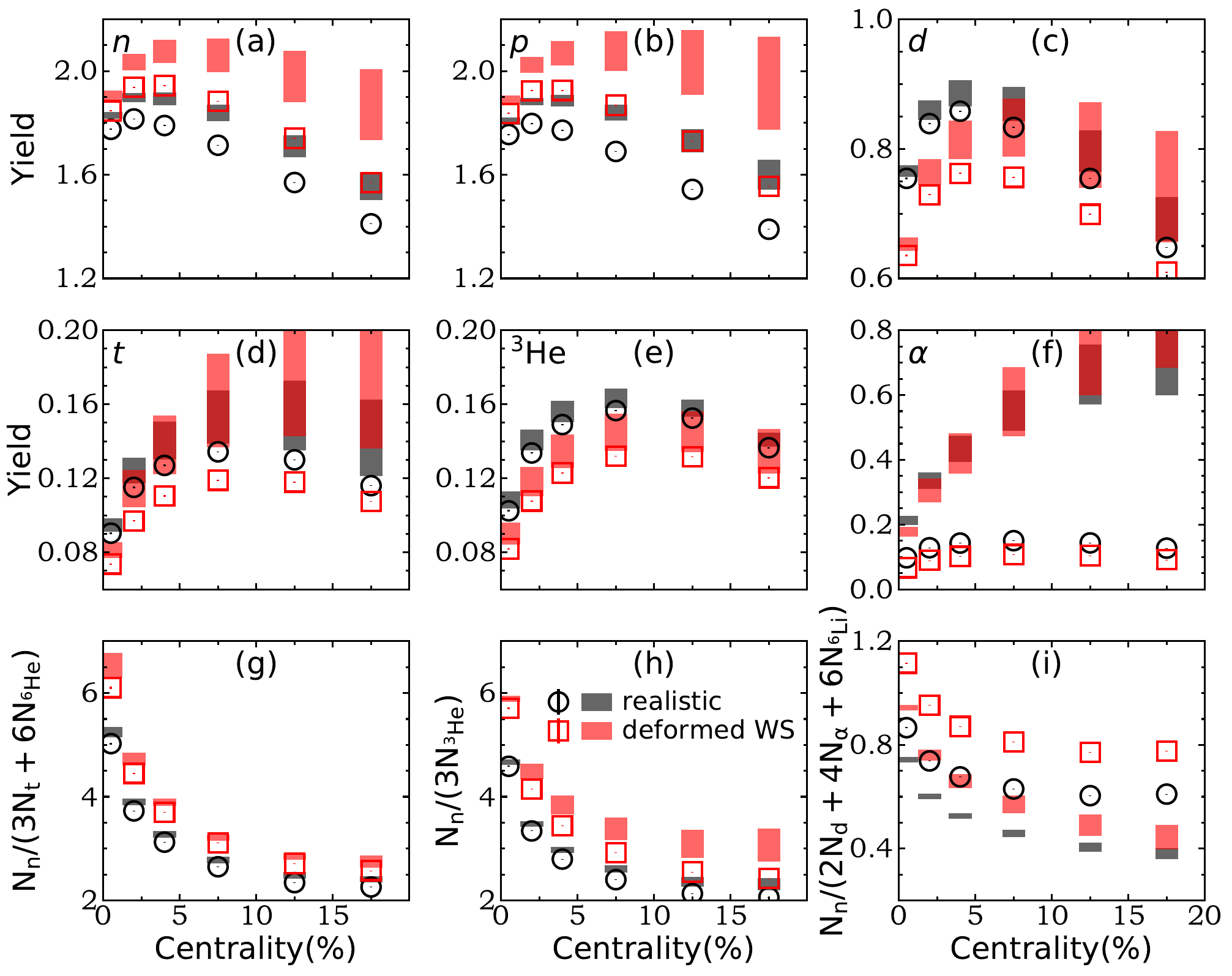}
\vspace*{-.5cm}
\caption{Yield of spectator neutrons (a), protons (b), deuterons (c), tritons (d), \(^3\mathrm{He}\) (e), and \(\alpha\) particles (f) as well as the corresponding yield ratios of free spectator neutrons to charged particles with mass-to-charge ratio \(A/Z = 3\) (g), \(3/2\) (h), and \(2\) (i) scaled by their constituent nucleon numbers, at \(0-20\%\) centrality in \(^{20}\)Ne + \(^{20}\)Ne collisions at \(\sqrt{s_\mathrm{NN}} = 200\) GeV. The bars and symbols represent results with and without the deexcitation process, respectively.}\label{fig:200}
\end{figure}

\textbf{Spectator probes.}
Now we turn to forward- and backward-rapidity particle yield from the same AMPT+specCOAL+testP+GEMINI simulations. Yield of spectator neutrons (\(n\)), protons (\(p\)), deuterons (\(d\)), tritons (\(t\)), \(^3\mathrm{He}\), and \(\alpha\) particles at \(0-20\%\) centrality in \(^{20}\)Ne + \(^{20}\)Ne collisions at \(\sqrt{s_\mathrm{NN}} = 200\) GeV are shown in Fig.~\ref{fig:200} (a)-(f). The statistical errors are invisibly small compared to those in Fig.~\ref{fig:vndeltapt}. The bars and symbols represent results with and without the deexcitation of heavy clusters, respectively. The height of the bars indicates the uncertainty due to possible errors in evaluating heavy-cluster excitation energies, which is estimated to be \(\pm1\) MeV per nucleon, and this uncertainty is largely reduced in UCC. It is seen that the deexcitation process enhances significantly the yield of $\alpha$ particles due to their strong binding, compared to the case of other light nuclei. Regardless of whether the deexcitation process is included, the yield of free spectator nucleons generated in collisions of nuclei with \(\alpha\)-cluster structure (shown in Fig.~\ref{fig:densitydis} (b)) is generally lower than that in collisions of nuclei without \(\alpha\)-cluster structure (shown in Fig.~\ref{fig:densitydis} (c)). However, the existence of $\alpha$-cluster structure on the yield of light nuclei has an opposite effect. This can be attributed to more compact initial phase-space distribution of nucleons within $\alpha$ clusters, which produces less free nucleons but more clusters in the minimum spanning tree algorithm and the Wigner-function coalescence approach, suppressing the production of free nucleons by about $5\%$ but enhancing that of light nuclei by more than $15\%$ in UCC. The $^{16}$O+$\alpha$ bicluster configuration as proposed in Ref.~\cite{Yamaguchi:2023mya} could have the similar effect as the five-$\alpha$ configuration used in the present study.

Since the existence of $\alpha$-cluster structure has an opposite effect on the yields of free nucleons and light nuclei, we propose to use the ratio of free spectator neutron yield ($N_n$) to that of charged particles as a probe. This would effectively isolate the $\alpha$-cluster effect and reduce the theoretical uncertainty due to, e.g., the deexcitation process during which a considerable amount of particles (especially $\alpha$ particles) are produced. The reduction of the theoretical uncertainty is due to the enhanced (suppressed) production of both free nucleons and light nuclei from the deexcitation of heavy clusters if their excitation energies are overestimated (underestimated). Since light nuclei with the same mass-to-charge ratio are not distinguishable by the detector in the forward-rapidity region~\cite{Tarafdar:2014oua}, which can only measure the energy deposit but is unable to identify particle species, the yield ratio $\frac{N_n}{\sum_{i \in \{A/Z=2\}} A\times N_{i}}$ could be a useful probe. As shown in Fig.~\ref{fig:200} (i), we consider light nuclei with $A/Z=2$ up to $^6$Li, and the initial condition with $\alpha$-cluster structure leads to the reduction of this ratio by about $20\%$. This effect is seen to be even larger in \(^{20}\)Ne + \(^{20}\)Ne collisions compared to \(^{16}\)O + \(^{16}\)O collisions~\cite{Liu:2023gun}, due to an extra \(\alpha\) cluster in \(^{20}\)Ne besides those in the tetrahedral structure. On the other hand, it could be experimentally challenging to distinguish spectator particles with \(A/Z = 2\) from $^{20}$Ne beam remnants. In that case, the alternative probes could be $\frac{N_n}{\sum_{i \in \{A/Z=3\}} A\times N_{i}}$ and $\frac{N_n}{\sum_{i \in \{A/Z=3/2\}} A\times N_{i}}$. We consider spectator light nuclei with $A/Z=3$ up to $^6$He, which may arrive the detector before its weak decay from our estimate, and spectator light nuclei with $A/Z=3/2$ only for $^3$He since $^6$Be is unable to reach the detector before its strong decay. The $\alpha$-cluster effect leads to a reduction of about $19\%$ for $\frac{N_n}{\sum_{i \in \{A/Z=3\}} A\times N_{i}}$ and about $21\%$ for $\frac{N_n}{\sum_{i \in \{A/Z=3/2\}} A\times N_{i}}$, as shown in Fig.~\ref{fig:200} (g) and (h), respectively.







\begin{figure}[ht]
\includegraphics[width=1.\linewidth]{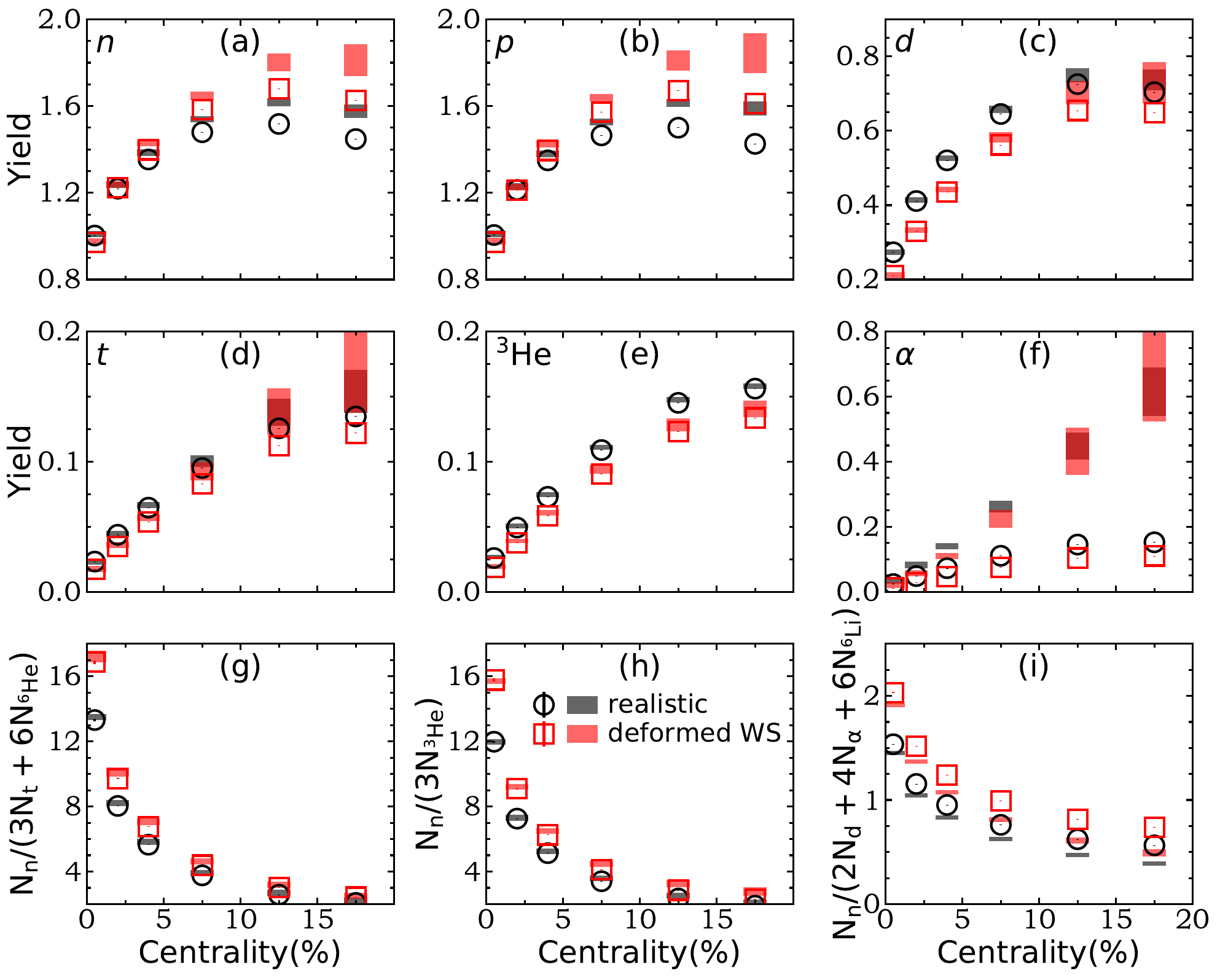}
\vspace*{-.5cm}
\caption{Similar to Fig.~\ref{fig:200} but for \nene\ collisions at \sqrtsnn = 7 TeV.}\label{fig:7000}
\end{figure}

The situation could be further improved at LHC energy, e.g., at \(\sqrt{s_\mathrm{NN}} = 7\) TeV as shown in Fig.~\ref{fig:7000}. Due to the larger nucleon-nucleon inelastic cross section at the higher collision energy, the number of spectator nucleons is largely reduced, leading to smaller yields of free spectator nucleons and light nuclei especially at small centralities. This also helps to reduce the uncertainty due to the deexcitation process as well. Therefore, the yield ratios $\frac{N_n}{\sum_{i} A\times N_{i}}$ with particle species $i$ for light nuclei of \(A/Z = 3\), \(3/2\), and \(2\) serve as better probes compared to those at RHIC energy. Quantitatively, the $\alpha$-cluster effect may reduce the yield ratios $\frac{N_n}{\sum_{i} A\times N_{i}}$ by about 25\% at \sqrtsnn = 7 TeV compared to about 20\% at \sqrtsnn = 200 GeV, in ultra-central \(^{20}\)Ne + \(^{20}\)Ne collisions.

\textbf{Summary and outlook.}
To summarize, we try to illustrate that relativistic heavy-ion collisions can not only probe the global shape of colliding nuclei but also probe their internal $\alpha$-cluster structure, and we take $^{20}$Ne proposed in Ref.~\cite{Jia:2022ozr} and investigated in Refs.~\cite{Giacalone:2024ixe,Giacalone:2024luz} as an example. Based on a hybrid simulation framework by combining the AMPT model with the specCOAL+TestP+GEMINI spectator particle generator, we have explored the possibility of distinguishing the density distributions of $^{20}$Ne with and without internal $\alpha$-cluster structure via mid-rapidity observables and free spectator particle yields in relativistic \nene\ collisions. Since the two density distributions give the same nucleus size and deformation, it is difficult to distinguish them with mid-rapidity observables. However, a considerable suppression of free spectator nucleon yield together with an appreciable enhancement of spectator light nuclei yield is observed from the initial condition with $\alpha$-cluster structure, due to the more compact initial phase-space distribution of nucleons inside pre-existing $\alpha$ clusters, compared to that from the density distribution without $\alpha$-cluster structure. In order to isolate the $\alpha$-cluster effect and reduce the uncertainty of the deexcitation process, we have further proposed that the yield ratios of free spectator neutrons to charged particles with \(A/Z = 3\), \(3/2\), and \(2\), scaled by their constituent nucleon numbers, serve as sensitive probes of \(\alpha\)-cluster structure in \(^{20}\)Ne. It has been illustrated that these yield ratios are reduced by about 25\% at $\sqrt{s_\mathrm{NN}} = 7$ TeV and about 20\% at $\sqrt{s_\mathrm{NN}} =200$ GeV in ultra-central \(^{20}\)Ne + \(^{20}\)Ne collisions with $\alpha$-cluster structure. For the first time, our study has proposed a direct probe for the existence of internal $\alpha$-cluster structure in $^{20}$Ne with relativistic heavy-ion collisions and free from the uncertainty of mid-rapidity dynamics.

\textbf{Acknowledgments.}
We acknowledge helpful discussions with Wen-Hao Zhou and Fu Ma as well as valuable comments from Rui Wang and Yi-Feng Sun. This work is supported by the National Key Research and Development Program of China under Grant Nos. 2022YFA1602404, 2022YFA1604900, and 2024YFA1612600, the Strategic Priority Research Program of the Chinese Academy of Sciences under Grant No. XDB34030000, the National Natural Science Foundation of China under Grant Nos. 12375125, 12225502, 12147101, 12075061, 12035011, and 11975167, the China Postdoctoral Science Foundation under Grant No. 2024M750489, the Natural Science Foundation of Shanghai under Grant No. 23JC1400200, the Shanghai Pujiang Talents Program under Grant No. 24PJA009, DOE Research Grant No. DE-SC0024602, and the Fundamental Research Funds for the Central Universities.

\bibliography{ref}
\end{document}